\documentclass[12pt,preprint]{aastex}
\usepackage{epsfig}

\shorttitle{Gemini Spectroscopy of UY Vol}
\shortauthors{Mikles et al.}

\begin{document}

\title{Gemini/GMOS Spectroscopy of  EXO 0748-676 (=UY Vol) in Outburst}
\author{Valerie J. Mikles}
\affil{Department of Physics \& Astronomy, Louisiana State University, Baton Rouge, LA 70803}
\email{vmikles@phys.lsu.edu}

\and

\author{Robert I. Hynes}
\affil{Department of Physics \& Astronomy, Louisiana State University, Baton Rouge, LA 70803}
\email{rih@phys.lsu.edu}

\begin{abstract}
  We present a phase-resolved, optical, spectroscopic study of the
  eclipsing low-mass X-ray binary, EXO 0748-676 = UY Vol. The
  sensitivity of Gemini combined with our complete phase coverage
  makes for the most detailed blue spectroscopic study of this source
  obtained during its extended twenty-four year period of activity. We
  identify 12 optical emission lines and present trailed spectra,
  tomograms, and the first modulation maps of this source in
  outburst. The strongest line emission originates downstream of the
  stream-impact point, and this component is quite variable from
  night-to-night.  Underlying this is weaker, more stable axisymmetric
  emission from the accretion disk.  We identify weak, sharp emission
  components moving in phase with the donor star, from which we
  measure $K_{\rm em} = 329\pm26$\,km\,s$^{-1}$.  Combining all the
  available dynamical constraints on the motion of the donor star with
  our observed accretion disk velocities we favor a neutron star mass
  close to canonical ($M_1\simeq 1.5$\,M$_{\odot}$) and a very low
  mass donor ($M_2 \simeq 0.1$\,M$_{\odot}$).  We note that there is
  no evidence for CNO processing that is often associated with
  undermassive donor stars, however.  A main sequence donor would
  require both a neutron star more massive than 2\,M$_{\odot}$ and
  substantially sub-Keplerian disk emission.
\end{abstract}


\section{Introduction}

The structure and stability of compact stellar objects depends on the
composition and equation of state (EOS) of their matter. Degenerate
white dwarf matter consists of protons and electrons and can be
described employing only special relativity, but in a neutron star,
the nature of matter has changed and general relativistic effects are
too large to be ignored, making the neutron star EOS a strong test of
General Relativity. The EOS determines both the maximum mass and the
mass-radius relation of a neutron star \citep{lattimer07}.

The majority of neutron star masses measured in binary pulsars are
consistent with the canonical value of $1.35 \pm 0.04 M_\odot$
\citep{thorsett99}, however, low-mass X-ray binaries (LMXBs) have
revealed a broader mass spread since accreting neutron stars grow
beyond their birth masses. This is consistent with higher masses
deduced in millisecond pulsars \citep[e.g.,][]{demorest10}, which
evolve from LMXBs. In order to discriminate between theoretical models
and discern the appropriate EOS, it is important to accurately
determine the mass distribution and the high-mass cutoff for neutron
star stability.

Because LMXBs are rich in emission lines, phase-resolved spectroscopy
can reveal the mass and geometry of the system.  Determining neutron
star masses in LMXBs remains challenging, however, because most
neutron star LMXBs are persistent X-ray sources in which the companion
is never revealed directly. It has been shown in Sco X-1 and other
sources that the motion of companion stars is visible in fluorescent
nitrogen and carbon lines \citep[the `Bowen blend' of N\,{\sc iii} and
C\,{\sc iii} around 4640\AA;][]{mcclintock75}. Several systems have
shown sharp emission lines tracing the motion of the companion
\citep{steeghs02,casares03,hynes03,casares06,cornelisse07a,cornelisse07,cornelisse07c}. The
ability to detect these narrow lines is dependent on the fraction of
emission line light originating from the companion relative to that
from the disk.

Although knowledge of the radial velocity curve of the companion star
is required for a complete determination of the system parameters,
much can also be learned about the accretion disk by studying trailed
spectra, Doppler tomograms, and modulation maps. Doppler tomography
uses kinematic information from lines to create a velocity map of a
system, even unraveling the complex information in blended lines
\citep{marsh05}. Modulation mapping is an extension of Doppler
tomography that permits mapping of time-dependent emission sources
\citep{steeghs03}. The analysis relaxes a fundamental constraint of
Doppler tomography by allowing emission line flux to vary, then
generates three tomograms simultaneously mapping the modulated and
non-modulated emission.

Among neutron stars, EXO 0748-676 (=UY Vol) has held particular
promise for constraining the EOS of dense nuclear matter. UY Vol was
first recognized by \citet{parmar85} as a transient source, and then
persisted in outburst for over twenty years before returning to
quiescence \citep{wolff08b,wolff08a, hynes08, torres08}. First
identified with EXOSAT, UY~Vol was later classified as a low-mass
X-ray binary (LMXB) containing a neutron star because it shows Type I
X-ray bursts \citep{gottwald86}. The source shows irregular X-ray dips
and periodic X-ray eclipses indicative of a 3.82 hour X-ray period
\citep{wolff2002}. The X-ray eclipses require a high inclination,
reducing a major source of uncertainty in the system parameter
estimates \citep{parmar86,hynes06}. The ephemeris of the source has
also been studied extensively, revealing complex period changes
\citep[see, e.g.,][]{wolff2002}, though since the source has recently
transitioned into quiescence \citep[$L_X \approx 8.5 \times 10^{33}
\rm{erg/s}$;][]{wolff08a}, it is no longer possible to perform
high-timing resolution analysis. There was considerable excitement
associated with the putative detection of gravitationally red-shifted
absorption lines from the neutron star surface \citep{cottam02}. This
detection could not be reproduced in a larger data set, however
\citep{cottam08}, and \citet{lin10} have argued that these lines
cannot originate from the neutron star surface in light of the now
known 552 Hz spin frequency of the neutron star.  Analyzing X-ray
burst emission from the source, \citet{ozel2006} estimated a neutron
star mass $M_1 > 1.8 M_\odot$ and possibly as high as $M_1 = 2.10 \pm
0.28 M_\odot$, although since this work used the gravitational
redshift, this too may need reexamination.

\citet{wade85} identified the optical counterpart UY Vol, which
brightened to $\rm{V} \sim$ 17 mag during the outburst, and faded to
$\rm{R} \sim$ 22 mag in 2009 \citep{hynes09}. \citet{pearson06}
performed Doppler tomography on 15 optical and ultraviolet emission
lines in UY~Vol. Their study combined observations from HST, VLT,
Magellan, CTIO 4m, and IUE. Their data are consistent with a 1.35
$M_\odot$ neutron star and main sequence companion, with a mass ratio
$q \approx 0.34$, based on the gas stream position shown in the
He\,{\sc ii} $\lambda$4686 \AA ~Doppler tomogram. \citet{munoz09} used
intermediate resolution spectroscopy to perform Doppler tomography on
narrow components in three He\,{\sc ii} emission lines. They measure
$K_{\rm em} = 300 \pm 10$ km/s, and derive a neutron star mass $M_1 >
1.5 M_\odot$ for the case of a main sequence companion. Modulation
mapping was successfully applied to the quiescent counterpart of UY
Vol by \citet{bassa09}. Analyzing the H$\alpha$ and He\,{\sc
  i}~$\lambda$6678\AA ~lines, they directly identify narrow line
emission from the counterpart and constrain $K_{\rm em} > 400$~km/s
from their tomograms.  They hence place a lower limit on the neutron
star mass of $M_1 > 1.27 M_\odot$ and estimate a mass ratio $0.075 < q
< 0.105$ for the specific case of a 1.4\,M$_{\odot}$ neutron star.
Most recently, \citet{Ratti:2012a} have performed further quiescent
observations and obtain $K_{\rm em}=308.5\pm3.9$\,km/s from H$\beta$ and
H$\gamma$ together with Fe\,{\sc ii}.  Surprisingly, if the widths of
their sharp components from the companion star are due to rotational
broadening, then $v \sin i = 255\pm22$\,km\,s$^{-1}$, and a neutron
star mass in excess of 3.5\,M$_{\odot}$ is required.  More likely is
that the widths are non-rotational and signatures of material
evaporating from the companion star in response to either X-ray
irradiation or a pulsar wind, making this a `Black-Widow-like' system.

In this paper, we present optical spectroscopy of UY~Vol obtained over
four nights in 2008, prior to the source's fade to quiescence. Our
observations provide complete phase coverage over multiple orbits. We
detect narrow emission components in the Bowen blend from the donor
star, and find a value of $K_{\rm em}$ consistent with the results of
\citet{munoz09}, however, we do not confirm their observation of
narrow He\,{\sc ii} emission lines originating near the donor
star. Our analysis represents the first attempt to deblend H\,{\sc
  i}/He\,{\sc ii} lines via Doppler mapping and also the first
application of modulation mapping of this source in outburst. In
Section 2, we describe our observations and data reduction. In Section
3 is our analysis and in Section 4, our discussion.

\section{Observations and Data Reduction}

We obtained intermediate dispersion, long slit spectroscopy with
Gemini-South using the Gemini Multi-Object Spectrograph
\citep[GMOS;][]{hook04} on 2008 February 9-11 and March 14. The
observations were designed to optimize spectral resolution for the
observation of narrow emission components and to get phase coverage
over multiple orbits. We obtained a total of 68 spectra each with 600s
exposure, totaling 11.3 hours of time ($> 3$ binary orbits). With a
1.0 arcsec slit and B1200 grating, we obtain a resolution of 30 km/s
at 4630 \AA. The spectra were reduced using the standard GMOS spectral
reduction packages in IRAF \footnote[1]{IRAF is distributed by the
  National Optical Astronomy Observatories, which are operated by the
  Association of Universities for Research in Astronomy, Inc., under
  cooperative agreement with the National Science Foundation.}. Over
the four nights, we had an average seeing of 0.85 arcsec and a maximum
seeing of 1.4 arcsec. Slit losses averaged 10\%. Dome flats were taken
nightly prior to and following each set of observations. The {\tt
  gemini.gmos} data reduction package in IRAF generates normalized
flat-fields, then applies bias subtraction, flat-fielding, and rough
wavelength calibration using archival arcs. Wavelength calibrations
are further refined using contemporaneous CuAr arc images taken during
the daytime. Gemini gratings are calibrated daily, so the zero-points
are well-known. Because Gemini has exceptional stability, the
pixel-to-wavelength transformations are not affected by significant
flexure and we did not require night-time arcs. The wavelength
calibrated image is then sky subtracted, and spectra are extracted
using {\tt gsextract}. Apertures were determined individually for each
image, and variance weighting is applied. To compute the phase for
each spectrum, we use the ephemeris given by \citet{wolff2002}. Our
observations and phase coverage are listed in Table \ref{tbl-1}.

We continuum fit each spectrum and generate both continuum normalized
and continuum subtracted spectra. We analyze both sets of spectra and
verify that our choice of continuum treatment has negligible effect on
our results.

\section{Analysis}

\subsection{The Average Spectrum}

\begin{figure*}
\plotone{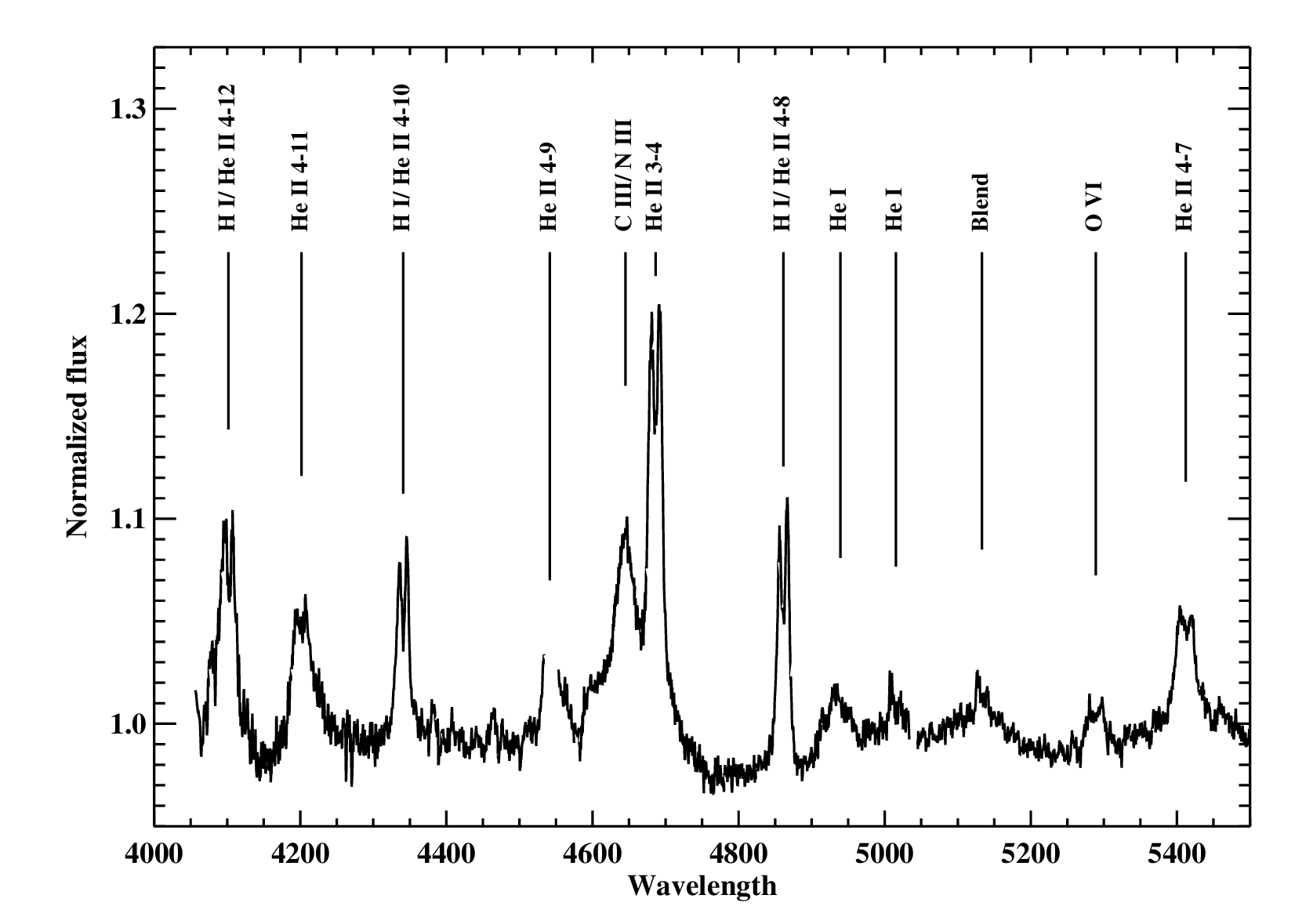}
\caption{The average, continuum normalized spectrum of UY Vol. We
  identify a variety of H/He lines with double-peak structures
  indicative of emission originating in the accretion disk. The
  He\,{\sc ii} 4-9 falls on a chip gap. Measured line centers are
  listed in Table \ref{tbl-2}.}
\label{fig:avgspec}
\end{figure*}

Combining 68 spectra, we create a continuum normalized, average
spectrum shown in Figure \ref{fig:avgspec}. We identify a total of 12
emission lines, sampling a range of ionization stages and
excitations. These lines are listed in Table \ref{tbl-2} with both
their laboratory wavelengths\footnote[2]{The Atomic Line List v2.04
  {\tt http://www.pa.uky.edu/~peter/atomic/}} and measured line
centers.

Both H\,{\sc i} and He\,{\sc ii} show double-peaked emission
characteristic of orbital motion. To measure the observed line
centers, we fit a double Gaussian to each line and determine the line
center of each peak, then average those values together. By dividing
our sample of 68 spectra in half, we create two combined spectra,
giving us two statistically independent measurements of the line
center so that we can constrain our errors. The separation of each
peak from the central wavelength for H\,{\sc i} and He\,{\sc ii} is
typically $\sim$ 402.8 $\pm$ 0.5 km/s.

We identify the Bowen blend in our spectrum. The Bowen blend is a
blend of C\,{\sc iii} and N\,{\sc iii} lines ranging from
$4638-4652$~\AA~\citep{mcclintock75}. In typical LMXBs, a relatively
equal proportion of C\,{\sc iii} and N\,{\sc iii} results in a blended
line observed at $4640$~\AA. Often times, strong narrow emission
features are observed \citep{cornelisse08}. In UY~Vol, the Bowen line
is centered at $4645$ \AA, suggesting a stronger contribution of
C\,{\sc iii} emission. GX~9+9 also shows this atypical ratio
\citep{cornelisse07}. We do not detect narrow line emission in our
individual spectra. This likely means that the disk dominates the
Bowen blend, but does not rule out weaker lines from the donor.

Two He\,{\sc i} lines $\lambda4922$, $\lambda5016$, and a blended
O\,{\sc vi} $\lambda 5289/5290$ are also detected. This O\,{\sc vi}
feature was previously identified by \citet{munoz09} as Fe\,{\sc ii}
$\lambda 5284$, but our trailed spectra (see below) clearly shows a
moving line centered at $\lambda 5289$. The identification of oxygen
at this ionization is consistent with \citet{pearson06} whose
ultraviolet spectrum showed O\,{\sc v} $\lambda 1371$ with very
similar kinematics. Other LMXBs also show O\,{\sc vi} line emission
\citep[see, e.g.][]{casares03,schmidtke00}. We identify no other Fe
lines in our spectrum of UY~Vol, so O\,{\sc vi} is a more plausible
interpretation.

\subsection{Trailed Spectra} 

\begin{figure*}
\includegraphics[width=\textwidth]{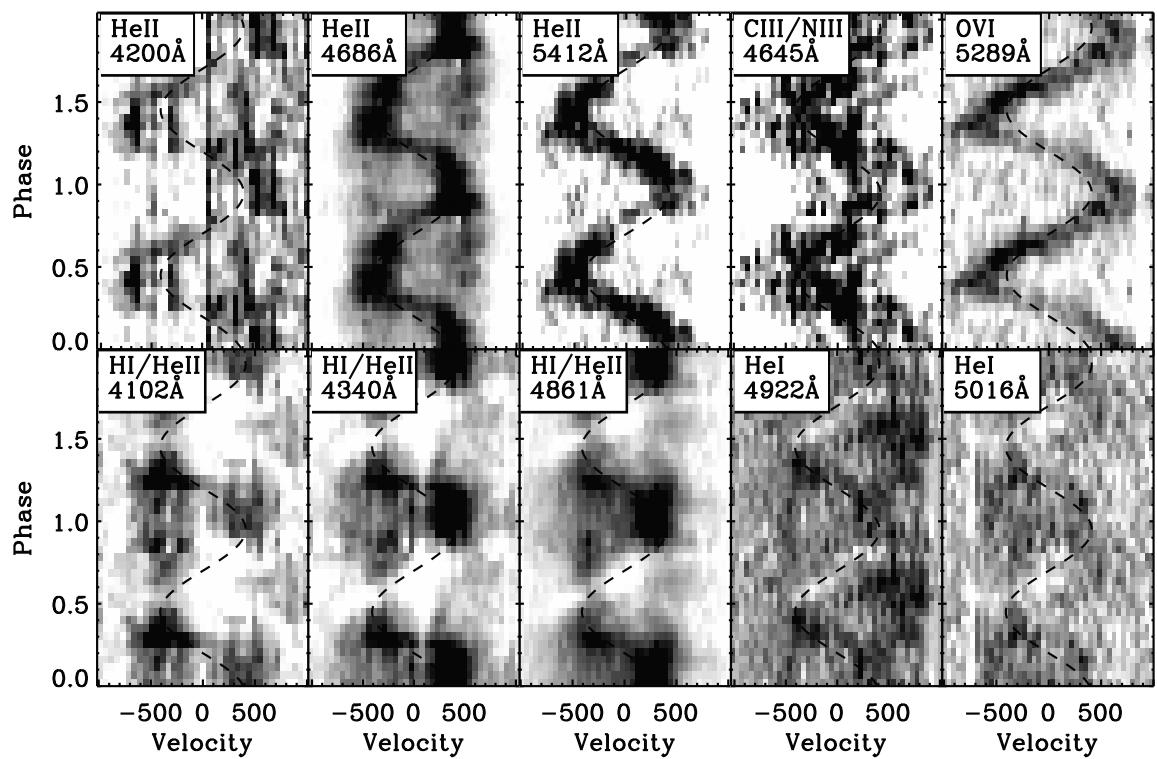}
\caption{Trailed spectra of ten prominent lines in UY Vol. The
  He\,{\sc ii} lines show clear S-wave patterns. The Bowen blend shows
  a broad S-wave feature with no narrow emission lines. The O\,{\sc
    vi} line shows a clear S-wave with an amplitude $K_{\rm
    em}=710$km/s. Trailed spectra for He\,{\sc i} and blended H\,{\sc
    i}/He\,{\sc ii} lines reveal phase-dependent absorption. Sine
  waves are over-plotted to guide the eye, with amplitude of 410 km/s
  and a phase offset of $\phi_0=0.3$. }
\label{fig:trailspec}
\end{figure*}

Binning our 68 spectra by phase, we create trailed spectra for each
observed line, and plot those in Figure \ref{fig:trailspec}. We
observe clear modulation of the line flux with phase in all lines, and
S-wave patterns emerge in several cases. The S-waves in the He\,{\sc
  ii} trailed spectra have a hooked shape with a sharp $dv/d\phi$ at
$\phi < 0.5$ and a smoother change for $\phi > 0.5$. Emission is also
relatively fainter at $\phi \sim 0.75$. The O\,{\sc vi} and Bowen
blend lines also show clear S-waves. The O\,{\sc vi} line shows higher
emission velocity than H and He (710 km/s vs. 403 km/s), indicating
either a location deeper inside the accretion disk or high velocity
disk overflow. Interestingly, in examining the LMXB 2A~1822--371,
\citet{casares03} find that the O\,{\sc vi} $\lambda$3811 emission
they observe has a radial velocity amplitude significantly larger than
that of H/He components.

Both He\,{\sc i} and H\,{\sc i}/He\,{\sc ii} blends show
characteristic absorption in the trailed spectra making it difficult
to identify a clear S-wave pattern. In He\,{\sc i} lines, absorption
occurs consistently at $0.7 < \phi < 0.9$. This is the same phase
range at which X-ray dips are seen, suggesting the absorption may also
be associated with the stream and its impact point. All lines showing
S-waves have phasing consistent with emission originating in the
accretion disk, either near the matter stream impact point or
downstream of it.

\subsection{The Doppler Corrected Spectrum}

\citet{munoz09} suggested that sharp components could be seen in the
Bowen line profile if it was Doppler-corrected to match a companion
star radial velocity semi-amplitude of $K_{\rm em}=300$~km/s, corresponding
to a feature seen in their He\,{\sc ii} Doppler tomogram.  We also
find that sharp components can be recovered by Doppler correction at
approximately this velocity, but have not been able to reproduce
corresponding features in a tomogram.  One should be cautious in
interpreting results of Doppler correction in isolation, as there
could be multiple spurious combinations of phase and radial velocity
that will produce small apparent peaks out of the noise.

 We Doppler correct using the equation:
\begin{equation}
V(\phi) = \gamma + K_{\rm em} \sin 2\pi(\phi - \phi_0).
\end{equation}
To quantify the Doppler correction procedure more rigorously, we
investigated the full parameter space in phase (0--1 in steps of 0.02)
and $K_{\rm em}$ (0--800\,km/s in steps of 10~km/s).  For each trial radial
velocity curve we Doppler correct the Bowen profiles, and then
cross-correlate them with a template consisting of one Gaussian
matching the instrumental resolution for each N\,{\sc iii} and C\,{\sc
  iii} component.  The individual strengths of the Gaussians were set
to match those listed by \citet{mcclintock75} with the overall
strength of nitrogen and carbon emission adjusted to best match our
data.  Using a cross-correlation in this way should favor finding
specific patterns of lines matching the Bowen blend wavelengths rather
than random peaks in the data, and should also result in a single
dominant peak for a good match, rather than the multiple peaks present
in the corrected line profiles.

\begin{figure*}
\plotone{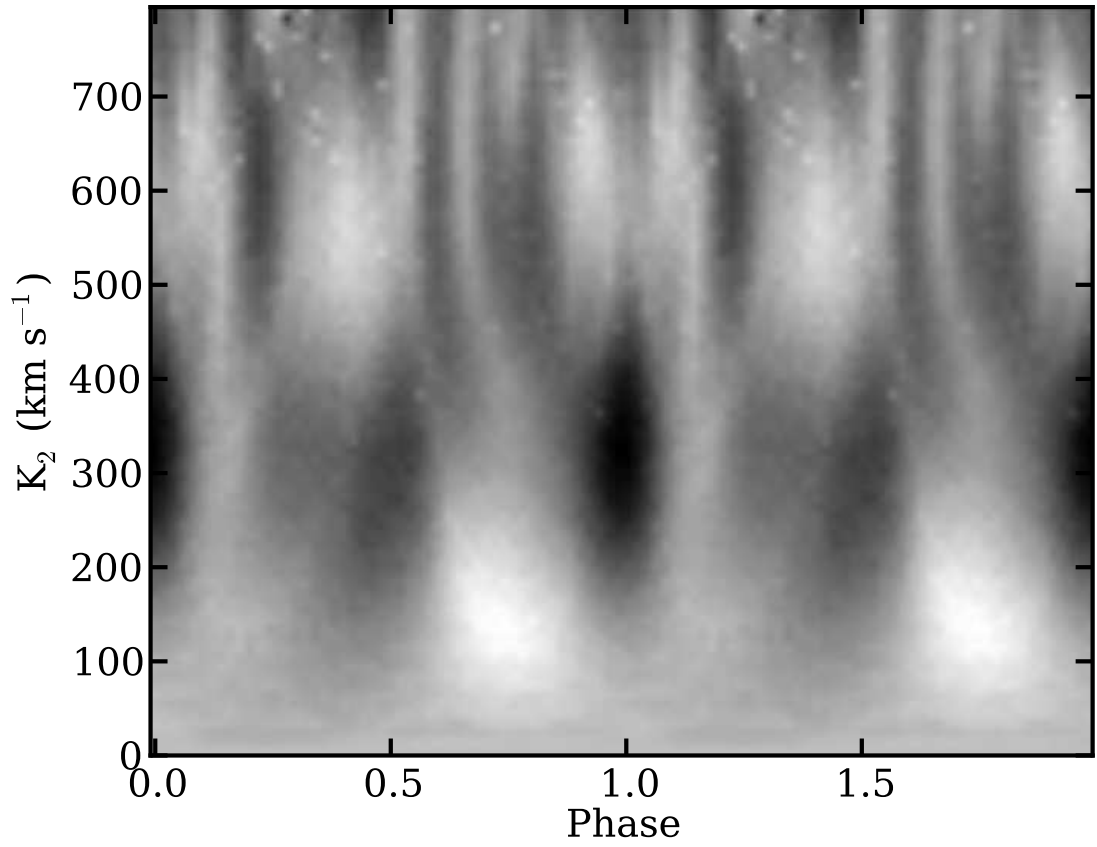}
\caption{The strength of the residual cross-correlation peak after
  subtraction of the broad component of the correlation corresponding
  to the disk.  A phase of 0 or 1 corresponds to a radial velocity
  curves in phase with the motion of the donor star.}
\label{DopCorrCCFFig}
\end{figure*}

After this procedure we find that the dominant feature in all of the
cross-correlation functions is a broad peak corresponding to the broad
emission associated with the disk.  For phases near to zero and
$K_{\rm em}$
near 300~km/s, a sharp peak is superposed on the broad maximum.  To
quantify the strength of this fit, we subtract a Gaussian fit to the
broad maximum and then measure the height of the residual central
peak.  We show the strength of this peak as a function of the trial
phases and $K_{\rm em}$ values in Figure \ref{DopCorrCCFFig}.

\begin{figure*}
\plotone{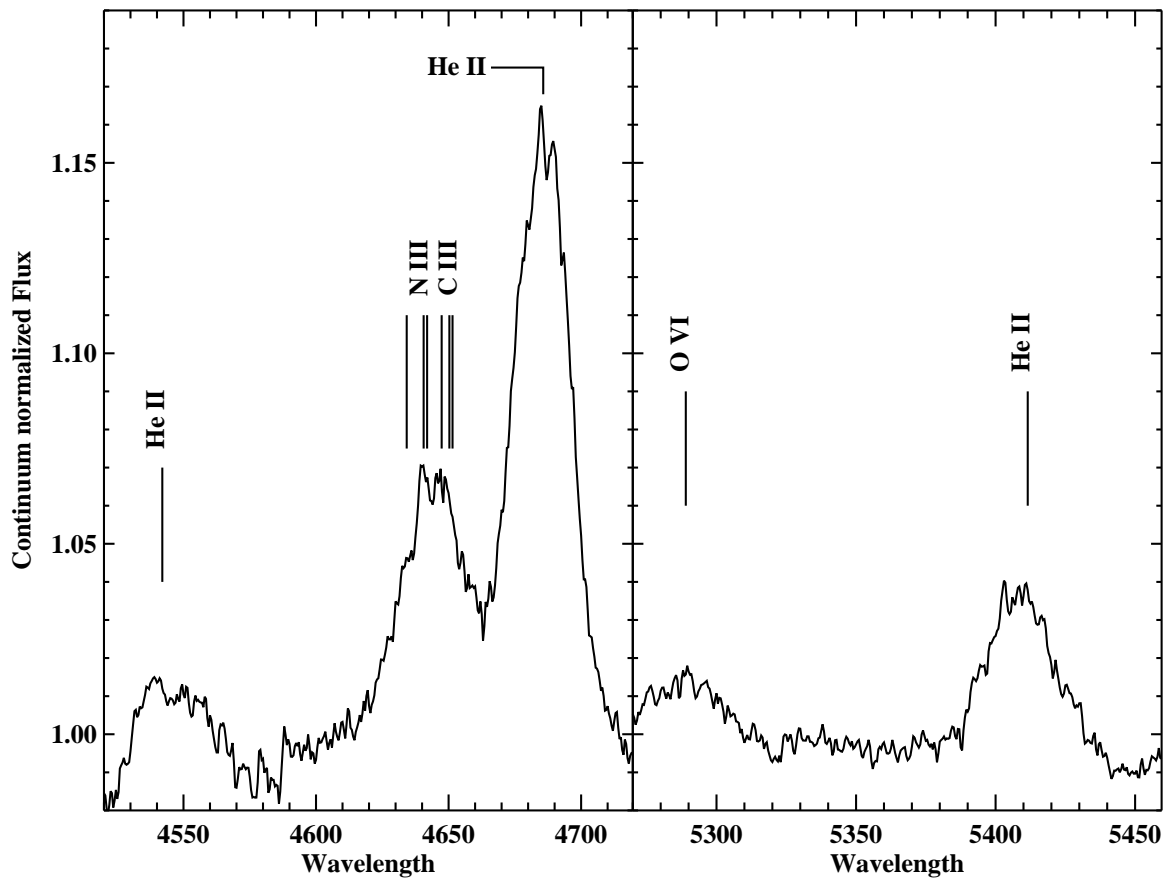}
\caption{We create a Doppler corrected spectrum using $K_{\rm
    em}=329$~km/s and $\gamma=70$~km/s. While we may tentatively
  identify narrow features in the Bowen blend, we do not observe
  strong narrow features in He\,{\sc ii}. Note that He\,{\sc ii}
  $\lambda$4542 occurs on a chip gap, so our Doppler reconstruction of
  that line is incomplete.}
\label{fig:sp_vcorr}
\end{figure*}

We see that the strongest feature is seen in phase with the motion of
the donor star. The peak is at a phase of $0.98$. We estimate $K_{\rm
  em} = 329\pm26$~km/s with the 1\,$\sigma$ uncertainty derived from a
bootstrap Monte Carlo simulation of our data analysis procedure.  The
velocity found is consistent with that of \citet{munoz09} and the
values determined from the quiescent He\,{\sc i} tomograms of
\citet{bassa09} and \citet{Ratti:2012a}. In Figure~\ref{fig:sp_vcorr},
we show the Doppler corrected spectrum of UY Vol around the
$\lambda$4686 and 5412 He\,{\sc ii} lines using the value $K_{\rm em}
= 329$~km/s and $\gamma = 70$~km/s.  We adopt the latter throughout
this work for consistency with \citep{munoz09}; we find a consistent
value from our O\,{\sc vi} line.

While the Doppler corrected spectrum, shown in Figure
\ref{fig:sp_vcorr} shows reasonable centering around the line centers,
we do not see the strong narrow line emission in He\,{\sc ii}. We
observe at most a very small contribution to the He\,{\sc ii} line
from the companion star unlike \citet{munoz09}. \citet{cornelisse08}
summarize a survey of 10 LMXBs, all of which show He\,{\sc ii}
emission that is not localized. However, many of these sources show
H\,{\sc i} in absorption. Both GX~9+9 and 2A~1822-371 show average
spectra with H\,{\sc i} in absorption
\citep{casares03,cornelisse07,cornelisse08}. All of our blended H/He
lines show a dual-peaked shape when Doppler corrected, with one peak
strongly centered near the He\,{\sc ii} emission and decreased
emission near H\,{\sc i}.

\subsection{Tomograms}

We next utilize the continuum subtracted spectra to perform Doppler
tomography, first rebinning the data to a constant velocity scale of
30 km/s/pix. We then create tomograms using {\tt doppler}
\citep{marsh05} which utilizes a combined maximum entropy and
chi-square minimization technique. In all cases, the entropy is
measured relative to a Gaussian blur default, meaning entropy is
insensitive to large scale variations. We generate tomograms extending
$\pm 1000$ km/s around the line center.

\begin{figure*}
\plotone{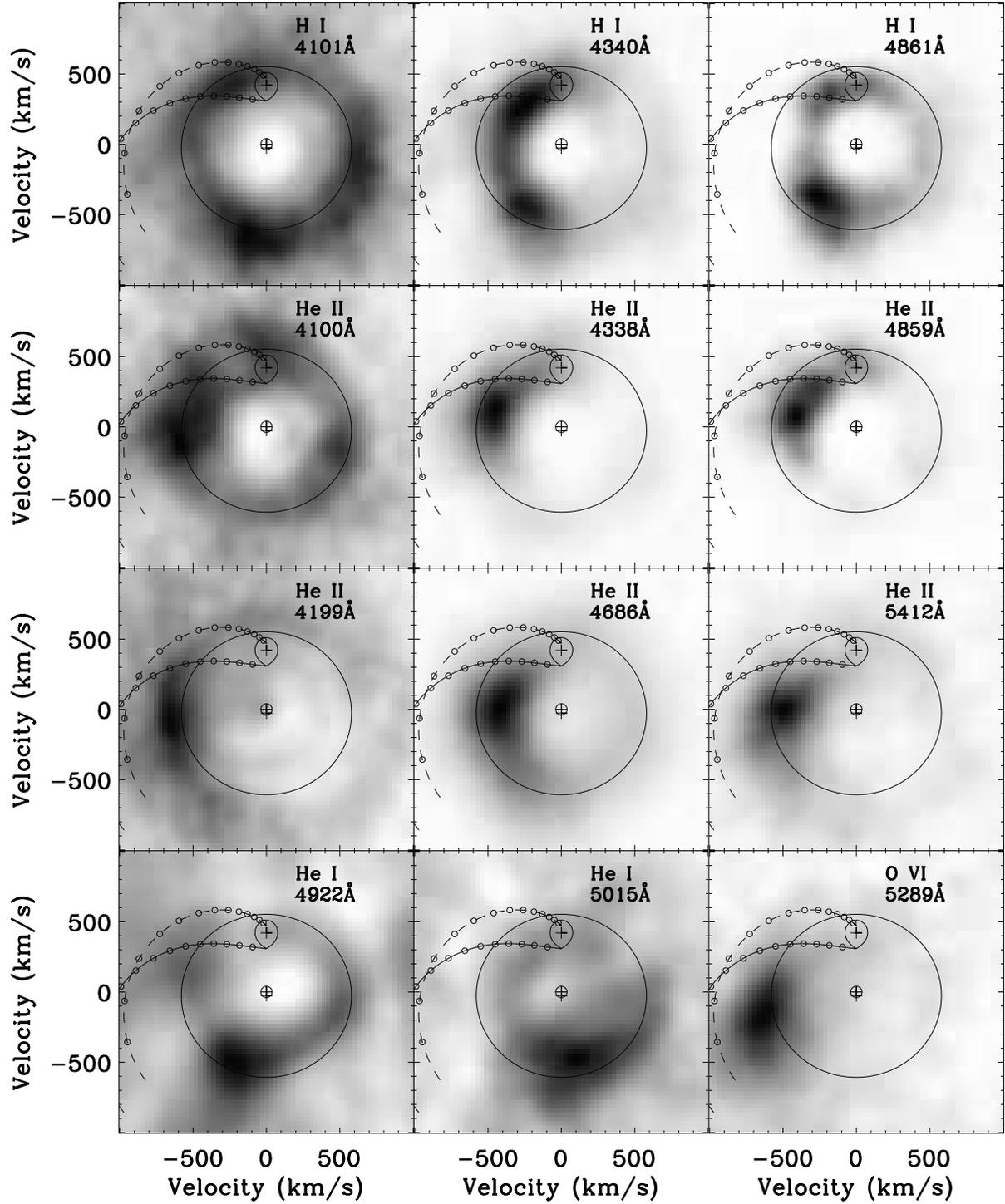}
\caption{Doppler tomograms of identified lines. The components for the
  H\,{\sc i} and He\,{\sc ii} lines are deblended and shown
  separately.}
\label{fig:doppler}
\end{figure*}

We use only the 49 spectra for which the source is out of eclipse. The
maps, shown in Figure \ref{fig:doppler}, include model locations for
the primary, secondary, matter stream, and L1 point for a 1.5
$M_\odot$ neutron star and an 0.1\,M$_{\odot}$ donor (see
Section~\ref{ParamSection}). The bulk of the line emission emanates
from the accretion disk, especially downstream of the matter impact
point, with no definitive contribution from the donor star. We were
not able to reconstruct a tomogram of the Bowen blend. The weak narrow
line emission and the volume of contributing components resulted in
poor maps.

In the top two rows of Figure \ref{fig:doppler}, we show tomograms of
our three H\,{\sc i}/He\,{\sc ii} lines. We used {\tt doppler} to
deblend the H/He contributions, and find that while the H\,{\sc i}
contribution is distributed throughout the disk, the He\,{\sc ii}
contribution is concentrated just downstream of the matter impact
point. The location of the deblended He\,{\sc ii} emission is
consistent with that seen in the isolated He\,{\sc ii} lines.

Using VLT data taken in 2003, \citet{pearson06} find He\,{\sc ii}
emission in $\lambda$4686, 5412 is brightest significantly closer to
the matter stream and assumed location of the secondary. In both
lines, \citet{pearson06} observe concentrated emission with $V_Y >
0$~km/s. Our data very consistently shows $V_Y \sim 0$~km/s. Our data
do not show the matter stream emission observed by
\citet{munoz09}. Their data do show some emission features at negative
$V_Y$. Since their data were taken one month {\it prior} to ours, this
precludes the existence of a concentrated source of He\,{\sc ii}
emission migrating along the streaming disk.

\begin{figure*}
\plottwo{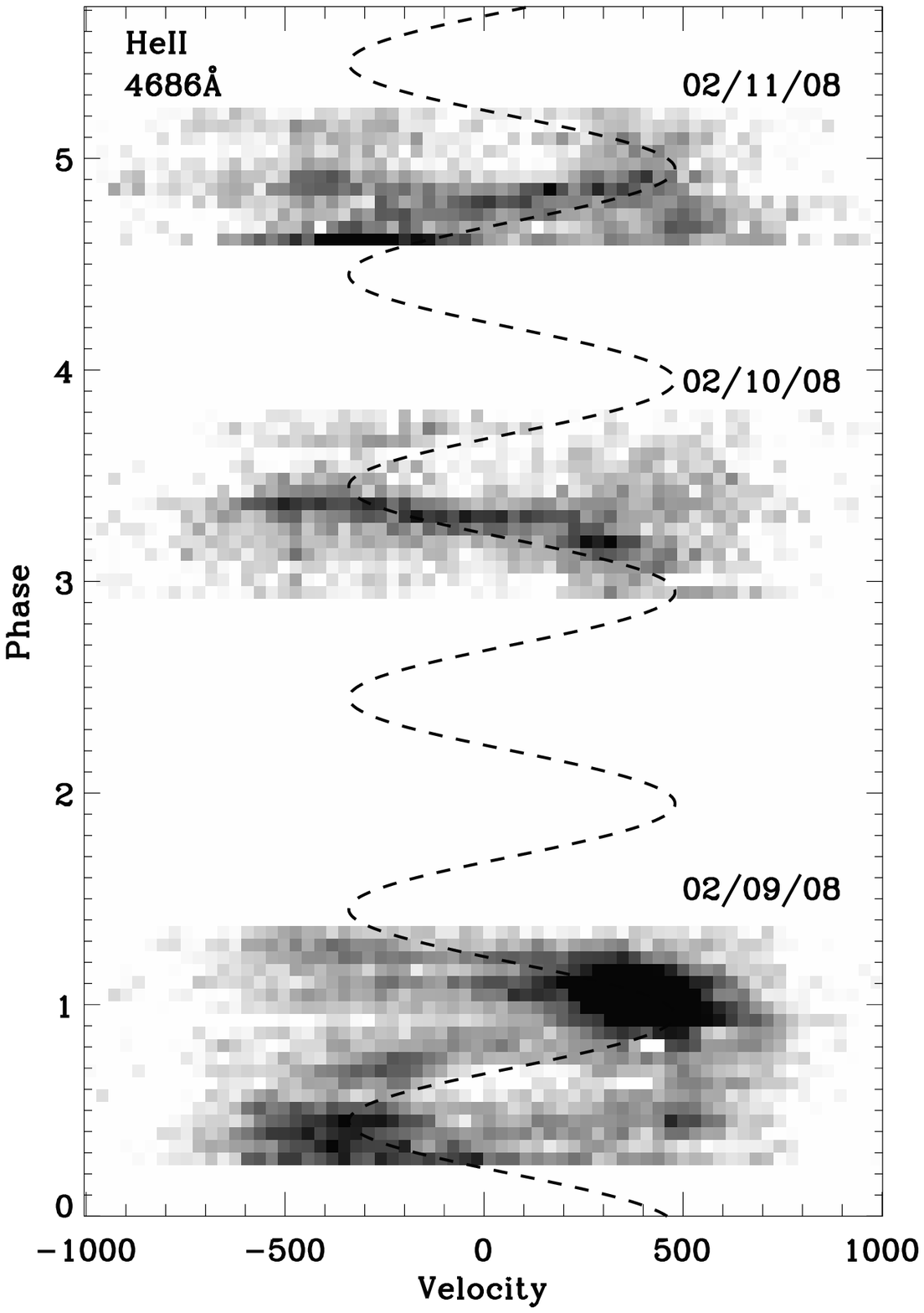}{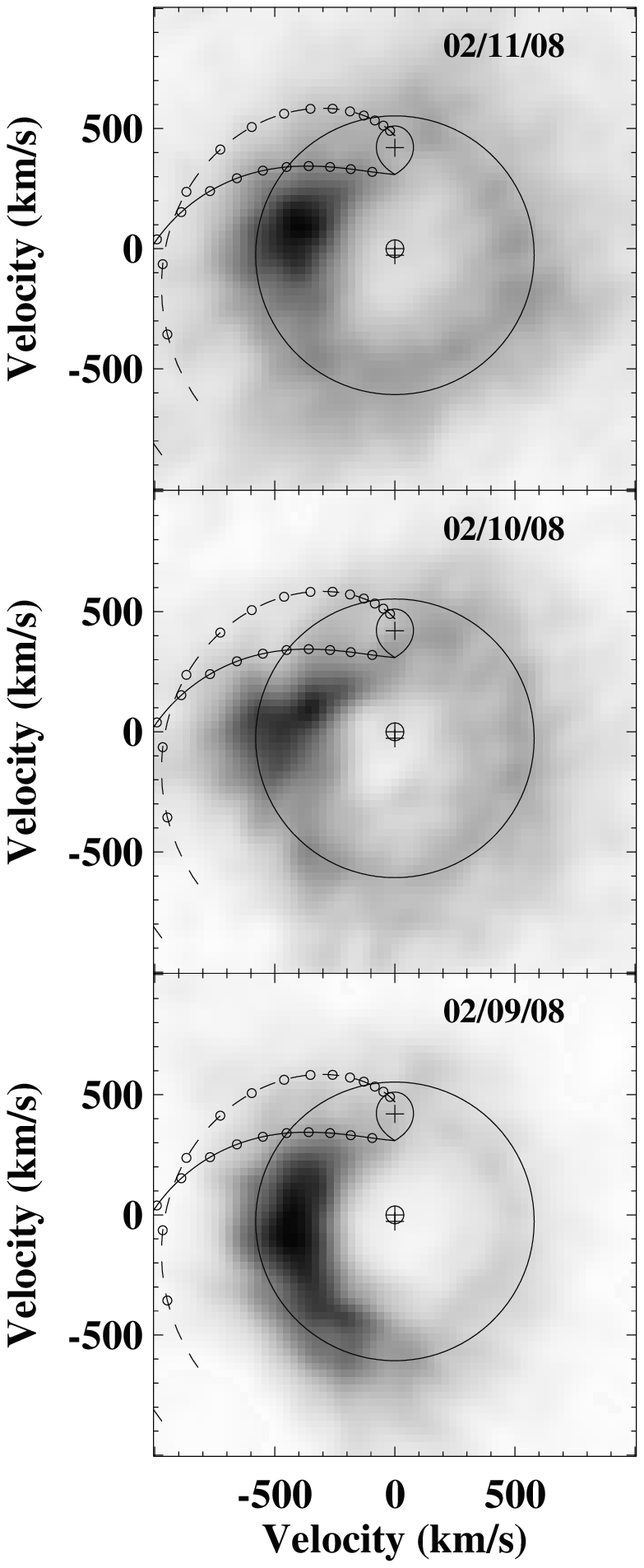}
\caption{Night-by-night tomograms of He\,{\sc ii} 4686.  Note, the
  tomograms for 2/10 and 2/11 are shown at a fractional gray-scale
  relative to the 2/9 for clarity. They only achieve 50\% of the
  signal. This is apparent by the relative peaks in the trailed
  spectra.}
\label{fig:doppbyday}
\end{figure*}

Because our data set covers multiple orbits, we were able to group our
spectra and characterize the changes in the tomogram on a timescale of
days. In Figure \ref{fig:doppbyday}, we show the trailed spectra and
tomograms for three consecutive days for which we had sufficient
orbital coverage. On February 9, the emission in the trailed spectra
is notably brighter and the peaks broader. In the tomogram, this
corresponds to emission concentrated just downstream of the matter
impact point with a large spread of emission along the lower left
tomogram. The next evening, the peaks in the trailed spectra are
notably narrower and the corresponding tomogram emission is more
compact. On February 10, the emission disperses toward the upper right
and on February 11, the emission is most concentrated. These
variations in morphology also appear in the modulation maps (see
below). Given how rapidly the centroid and morphology of He\,{\sc ii}
emission can change from epoch to epoch, it is unlikely that emission
features in the tomogram associated with stream-disk impact region can
be used to constrain the mass ratio reliably as attempted by
\citet{pearson06}.

\begin{figure*}
\plotone{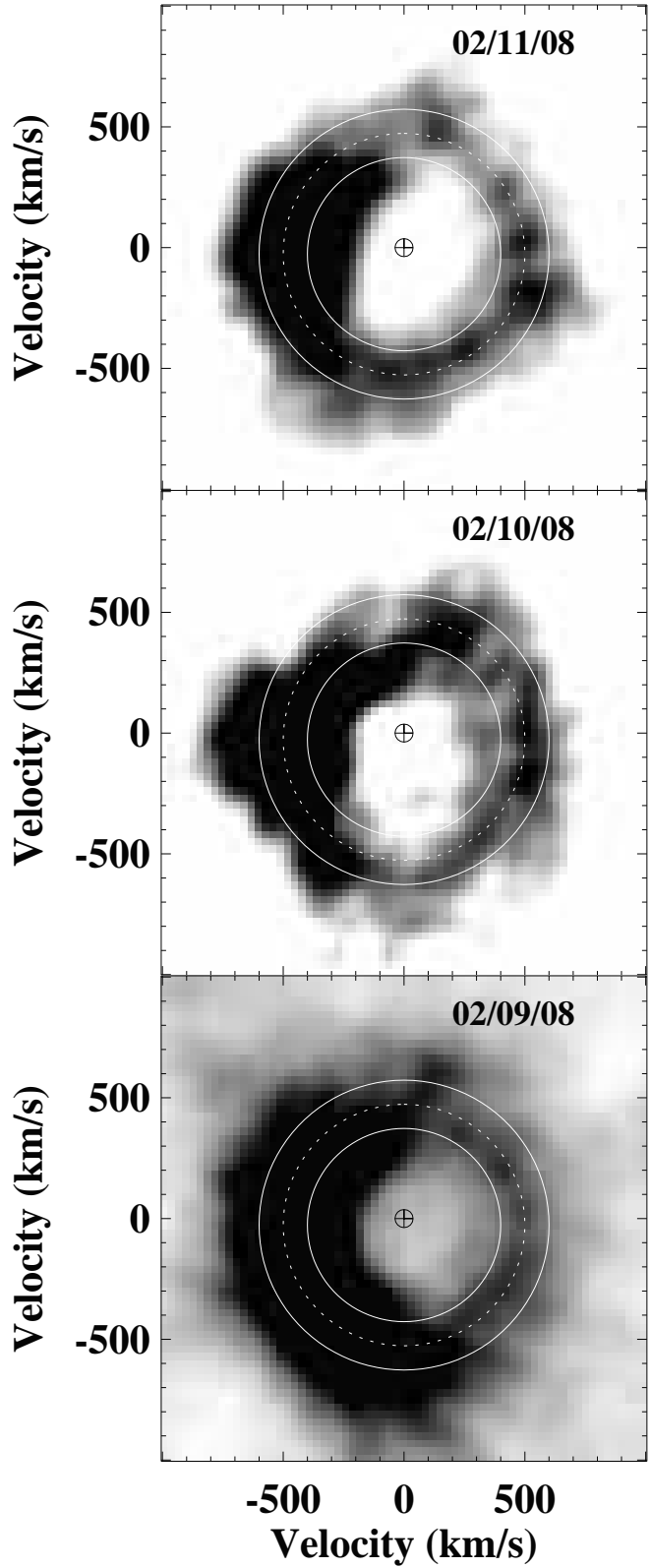}
\caption{Night-by-night tomograms of  He\,{\sc ii} 4686 as for
  Figure~\ref{fig:doppbyday}.  The gray-scale has now been adjusted
  individually to maximize visibility of the disk on the right hand
  side of the tomogram.  The plotted circles correspond to disk
  velocities of 400, 500, and 600\,km\,s$^{-1}$.}
\label{fig:doppbydaytwo}
\end{figure*}

Our He\,{\sc ii} 4686\,\AA\ maps do rather consistently define the
accretion disk emission, and this potentially provides an alternative
constraint on the system parameters.  We show the same night-by-night
tomograms in Figure~\ref{fig:doppbydaytwo}, with the gray-scale
adjusted to maximize visibility of the disk on the right hand side of
the tomogram.  We will concentrate on this region as it should not be
distorted by the changes seen in the stream-impact/bulge region that
vary from night-to-night.  Instead the kinematics of the right hand
side of the maps are very stable and define a circular ring quite
well.  The projected disk velocity indicated by the data is 500~km/s,
and is well bounded by rings at 400 and 600~km/s.  We will discuss the
implications of interpreting this as a Keplerian disk velocity in
Section~\ref{ParamSection}.

In the last row of Figure~\ref{fig:doppler}, we see that He\,{\sc i}
emits more strongly in a different region of the disk to He\,{\sc ii},
separated by about 0.25 in phase.  It is possible that the He\,{\sc
  ii} gets ionized near the matter impact point and is able to later
recombine and form He\,{\sc i} downstream.

The O\,{\sc vi} emission is fairly concentrated, suggesting
temperature conditions unique to O\,{\sc vi} excitation at this point
in the disk. \citet{pearson06} observe O\,{\sc v} emission in the
UV. As with our O\,{\sc vi} observations, their O\,{\sc v}
observations do not occur along the the ballistic stream or at
Keplerian velocity, however, it is consistent with a location
downstream of the impact or the stream overflow \citep{pearson06}. The
high velocity of the O\,{\sc vi} emission observed in the tomogram is
expected from the fact that the S-wave seen in our O\,{\sc vi} trailed
spectrum had a higher amplitude than that of the H/He lines.

The Doppler tomograms begin to form a consistent picture of the
ionization structure of the disk. He\,{\sc ii} is observed in the
ionized disk bulge, predominantly on the left of the tomogram at
negative $V_X$ and low $V_Y$. While we do observe some variance in the
central location, we can say that we do not observe He\,{\sc ii}
emission from the donor star. We also know that the prominent He\,{\sc
  ii} emission occurs downstream of the matter impact point.  He\,{\sc
  i} is absent from the ionized disk bulge. It occurs most prominently
in the lower part of the tomogram at negative $V_Y$, along the flow of
the disk stream. As it occurs on the opposite side of the disk from
the matter stream, it is probable that the temperature conditions in
the disk are more amicable to the recombination of He\,{\sc ii}. The
O\,{\sc vi} emission, in contrast, may be tracing a higher ionization
region inside the disk.

\subsection{Modulation Maps}
\begin{figure*}
\includegraphics[width=12cm,angle=0]{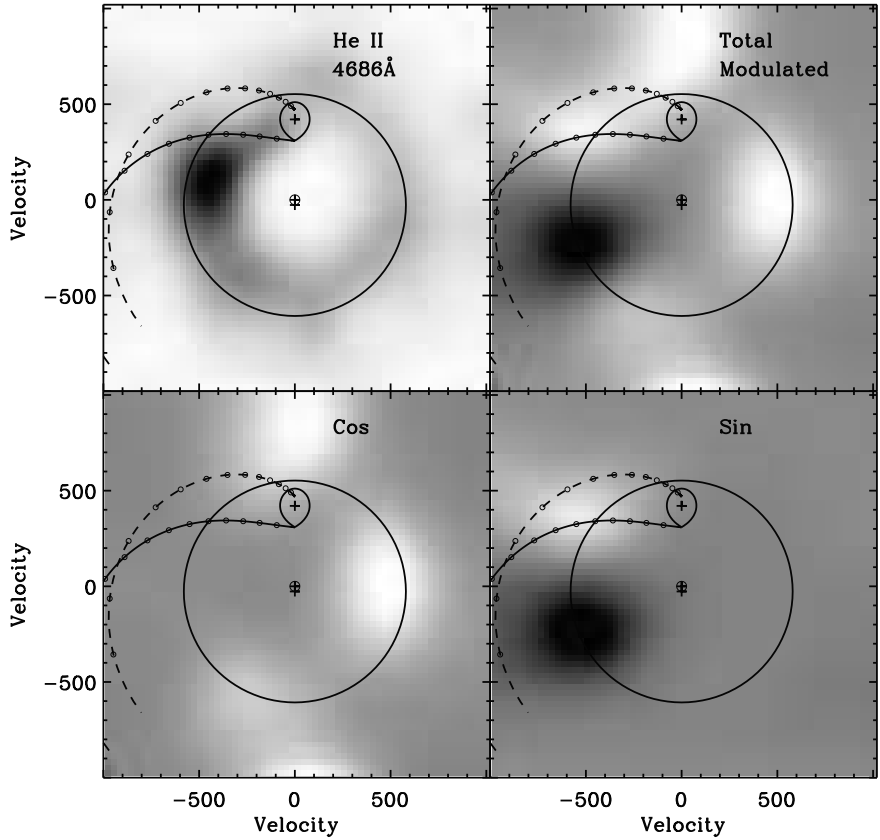}
\caption{Modulation maps of the He\,{\sc ii} $\lambda$4686
  line. Starting at the upper left and going clockwise, the panels are
  (1) the average non-modulating emission, (2) the total modulated
  emission ($I_{sin}+I_{cos}$), (3) the $I_{sin}$ component of the
  modulated emission, and (4) the $I_{cos}$ component of the modulated
  emission. The modulation maps are shown at a fractional gray-scale
  of $\pm 6$\% of the average, non-modulating emission.}
\label{fig:modmap4686}
\end{figure*}

\begin{figure*}
\includegraphics[width=12cm,angle=0]{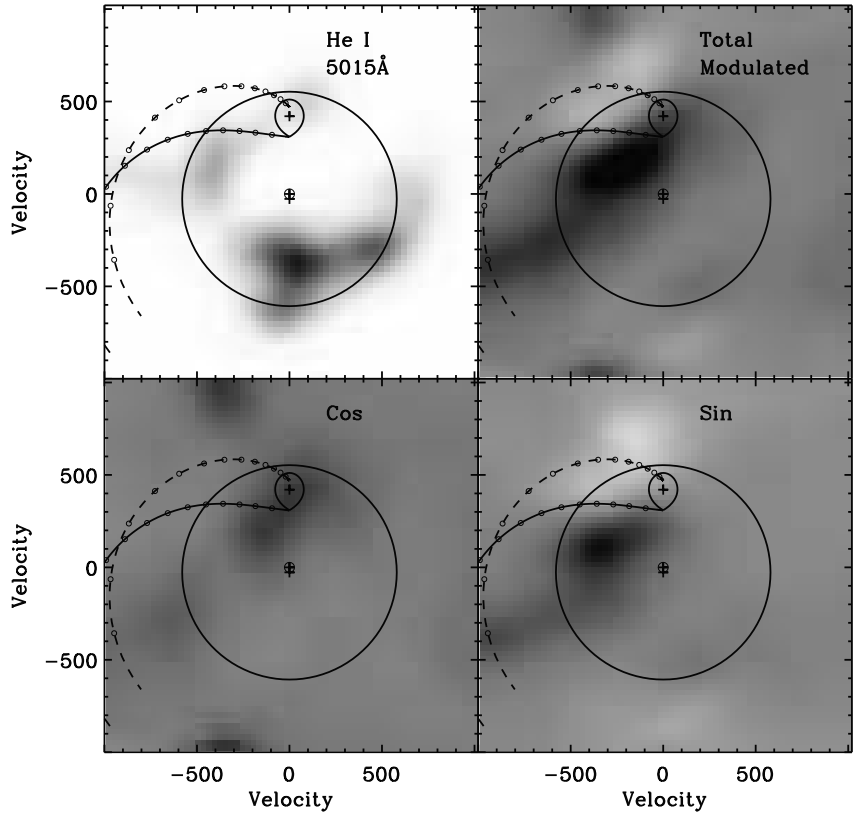}
\caption{Modulation maps of the He\,{\sc i} $\lambda$ 5015 line. The
  panels are as described in \ref{fig:modmap4686}. The fractional
  gray-scale employed here is $\pm 35$\% of the average non-modulating
  emission.}
\label{fig:modmap5015}
\end{figure*}

Modulation mapping is an extension of Doppler tomography that maps
sources varying harmonically as a function of the orbital period
\citep{steeghs03}.  Periodic absorption, disk flickering, and varying
visibility of the emission region all contribute to features in the
modulation map. Thus, we must go back to the trailed spectra for a
more accurate interpretation. At present, the modulation mapping code
does not account for eclipsing, so we use only 49 of the 68 spectra
for which the source was out of eclipse. Also, since the code does not
deblend lines, we focus our study on isolated lines. Although
modulation mapping works best with flux calibrated spectra, we did not
have a comparison star on the slit for relative flux calibration. We
assume that there is no significant continuum flux variation night to
night and create modulation maps of all identified lines and plot two
representative samples in Figures
\ref{fig:modmap4686}-\ref{fig:modmap5015}. Modulation mapping
considers a line source modulating harmonically with flux as a
function of orbital phase, such that
\begin{math}
f(\phi) = I_{avg} + I_{cos}\cos(2\pi \phi) + I_{sin} \sin(2\pi \phi).
\end{math}
 The modulation maps show the average, non-varying line flux
 ($I_{avg}$) as well as the total modulated emission
 $(I_{sin}+I_{cos}$), and the individual sine and cosine emission
 maps. Because the modulated emission is often small relative to the
 average emission, the maps are shown on a fractional gray-scale
 relative to the average emission. Note that $I_{sin}$ and $I_{cos}$
 can have positive or negative amplitudes and so their combination can
 model all phasings of the light curve.

 For the $\lambda4686$ He\,{\sc ii} line, the modulated emission
 contributes less than 6\% of the average emission. Figure
 \ref{fig:modmap4686} shows maps of the He\,{\sc ii} $\lambda4686$
 line. As with the Doppler tomogram, we observe the classic disk
 structure in emission, and concentrated emission just downstream of
 the stream impact point. Modulated emission is dominated by the sine
 term and appears most strongly downstream of the impact point. This
 corresponds to the region where the night-to-night tomograms differ
 most (see Fig. \ref{fig:doppbyday}). This modulated component would
 make our line emission appear brighter at $\phi \sim 0.25$ and
 fainter at $\phi \sim 0.75$. In the trailed spectrum, this is
 observed as a relative brightening in the S-wave near the peak radial
 velocity. A negative amplitude modulation appears in the cosine map,
 bordering the side of the disk farthest from impact. This would make
 our emission fainter at $\phi \sim 0$ and brighter at $\phi \sim
 0.5$. In the trailed spectrum, this appears as increased emission at
 $\phi \sim 0.5$ in anti-phase with the S-wave amplitude. The
 modulating component may correspond to the irradiated inner edge of
 the stream bulge.

 The split structure of the He\,{\sc i} $\lambda5015$ emission is
 readily apparent in the modulation map shown in Figure
 \ref{fig:modmap5015}. Modulated emission here contributes up to 35\%
 of the total flux. The sine term dominates the emission nearest the
 matter impact point. Very likely, this modulation does not represent
 varying visibility of an emission source, but rather the periodic
 absorption which is observed in the trailed spectra at $\phi \sim
 0.75$. The non-modulated emission, $I_{avg}$ is strongest on the far
 side of the disk, where He\,{\sc ii} is likely recombining.

 The presence of modulating emission contributes to our emerging
 picture of the disk. We certainly confirm the presence of ionized
 He\,{\sc ii} emission at or near the ionized disk bulge. The
 modulating feature in the sine map could represent the formation of
 optically thick clouds above and below the plane of the disk. These
 clouds are formed when the matter stream impacts the disk, and they
 are illuminated by X-rays from the disk. Since they are optically
 thick to X-rays, we would only observe these clouds at $\phi \sim
 0.25$. These clouds may also be responsible for the periodic
 absorption in He\,{\sc i} and H\,{\sc i} and would be responsible for
 irregular X-ray dipping discovered by \citet{parmar86}.

\section{Discussion}

\subsection{Accretion disk structure}

While Doppler tomography has long been used to interpret the accretion
disk structure, we must be cautious not to over interpret certain
aspects, as we have shown that prominent emission region morphology
can vary between orbits. Our tomograms of He\,{\sc ii} $\lambda$4686
and 5412 look different than those of \citet{pearson06} and
\citet{munoz09}, but this is likely reflective of the disk evolving
between epochs.

\citet{pearson06} attempted to constrain the mass ratio of UY~Vol by
studying emission from the matter stream as observed in their
tomograms for He\,{\sc ii} $\lambda$4686. Our observation of the same
line does not show prominent emission from the matter impact point,
but rather, some point on the disk downstream of the matter impact
point. At best, our H\,{\sc i} maps show some emission coincident with
the velocities expected for the matter stream, and He\,{\sc ii} is
observed further downstream on the left side of the disk. The presence
of He\,{\sc i} emission in the lower part of the disk suggests the
ionized He\,{\sc ii} is recombining as it moves around the disk.

Neither \citet{pearson06} nor this work confirm the concentrated
He\,{\sc ii} emission coincident with the location of the donor star
reported by \citet{munoz09}. While the morphology of prominent disk
emission is expected to change rapidly, emission from the donor star
should be more consistent. It is possible that the relative brightness
of narrow emission from the donor compared to the disk made it
unobservable at other epochs.  We note that the He\,{\sc ii} emission
associated with the donor star by \citet{munoz09} is part of a complex
extending to negative $V_x$ in their maps, and roughly parallelling
the stream trajectory.  It is then also possible the the He\,{\sc ii}
seen at $V_{x}=0$ by \citet{munoz09} originated from the beginning of
the matter stream rather than the donor star itself, although it is
unclear how He\,{\sc ii} emission is produced in this region that
should be shielded from direct irradiation by the disk rim.

As we must already use caution when comparing tomograms of the same
source from different epochs, so too we must cautiously approach how
we compare the tomograms of UY~Vol to other LMXBs. In examining the
source GX~9+9, \citet{cornelisse07} observe prominent He\,{\sc ii}
emission in the lower left quadrant of their tomograms, where UY~Vol
shows more prominent He\,{\sc i}. \citet{hynes01} find a similar
emission structure for He\,{\sc ii} in Doppler maps of
XTE~J2123-058. Both authors draw comparisons between these sources and
SW~Sex cataclysmic variables. The ionization structure of the emission
source suggests a matter stream not connected with the accretion disk,
that is either overflowing the disk or being propelled away
\citep[][and references therein]{hynes01,cornelisse07}.

In UY~Vol, the six He\,{\sc ii} lines for which we create tomograms
show emission most concentrated at $V_Y \sim 0$, but the $V_X$
velocities occasionally suggest that while most emission is consistent
with the Keplerian velocity of the outer edge of disk, some may be
sub-Keplerian representing a similar overflow.

\citet{casares03} performed a study of the LMXB 2A~1822-371. Their map
of He\,{\sc i} $\lambda$4471 places the He\,{\sc i} absorber on the
leading side of companion's Roche lobe or over the gas stream. It is
not surprising that their He\,{\sc i} maps are remarkably different
than ours since they observe He\,{\sc i} in absorption whereas UY~Vol
shows it in emission. Like UY~Vol, 2A~1822-371 shows oxygen
emission. \citet{casares03} observe O\,{\sc vi} 3811\,\AA\ clearly in
the post-shock region between the matter stream and the disk. This is
in a very different location than that observed for UY~Vol. Our
O\,{\sc vi} emission emanates downstream of the matter impact point,
at a high velocity, suggesting it occurs either deeper within the disk
or in a region of high-velocity disk overflow. It is worth nothing
that 2A~1822-371 is also at a higher inclination and likely higher
mass accretion rate than UY~Vol, thus the direct stream impact may be
more excited.

In summary, while UY~Vol does not show the abundance of He\,{\sc i}
absorbers seen in 2A~1822-371, it does seem to show similar ionized
emission in the disk and possible high-velocity and sub-Keplerian disk
overflows as seen in other LMXBs. Interestingly, in UY~Vol, we observe
He\,{\sc i} emission as part of the accretion disk structure, which
may give us some insight to the relative mass accretion rate and disk
temperature as compared to other LMXBs.

\subsection{Abundances}
\label{AbundanceSection}

We find that the C\,{\sc iii} emission in the Doppler corrected Bowen
blend profile in Figure~\ref{fig:sp_vcorr} is of comparable strength
to N\,{\sc iii}.  This can be compared to Sco~X-1 \citep{steeghs02},
2A~1822--371 \citep{casares03}, 4U~1636--536 \citep{casares06},
LMC~X-2 \citep{cornelisse07a}, and Aql~X-1 \citep{cornelisse07c} where
N\,{\sc iii} is dominant, 4U~1735---444, \citep{casares06} where
C\,{\sc iii} and N\,{\sc iii} are of comparable strength, and GX~9+9
\citep{cornelisse07} where C\,{\sc iii} dominates.  Thus the C\,{\sc
  iii}:N\,{\sc iii} ratio in UY~Vol is among the higher values seen
among neutron star LMXBs, suggesting that the carbon abundance is at
least normal, if not enhanced.  We should be wary of over-interpreting
the ratio, however, as it does depend on physical conditions as well
as abundances.  This is illustrated by the diversity of Bowen blend
structures seen in GX~339--4 \citep{hynes03}.

As discussed in the case of GX~9+9, the strong C\,{\sc iii} rules out
substantial CNO processing of the accreted material since this would
result in a pronounced deficit of carbon relative to nitrogen.  The
effects of CNO processing on LMXB spectra are even more pronounced in
the far-UV, where the normally dominant C\,{\sc iv} line can become
completely undetectable, as in the case of the black hole system
XTE~J1118+480 \citep{haswell02}.  In UY~Vol, on the other hand, the
far-UV spectrum is typical with a very strong C\,{\sc iv} line
\citep{pearson06}.

\subsection{System parameters}
\label{ParamSection}

There have now been four independent detections of features that
appear to be associated with the companion star in UY~Vol, spanning
four different emission lines and both outburst and quiescence.  In
outburst we find $K_{\rm em}=329\pm26$\,km\,s$^{-1}$ in the Bowen
blend, and \citet{munoz09} obtain $K_{\rm em}=300\pm10$\,km\,s$^{-1}$
from He\,{\sc ii}.  In quiescence \citet{bassa09} obtain
$345\pm5$\,km\,s$^{-1}$ in He\,{\sc i} and $410\pm5$\,km\,s$^{-1}$ in
H$\alpha$, while \citet{Ratti:2012a} find $308.5\pm3.4$\,km\,s$^{-1}$
from a weighted average of H$\beta$ and H$\gamma$.  There is some
spread in the deduced values, but there is also a striking congruence
between them with emission from the companion star apparently spanning
at least the range in velocity of 300--400\,km\,s$^{-1}$, with
different average velocities seen at different times and in different
emission lines.  This provides more information than a single
detection, as discussed by \citet{bassa09}, but also confuses
interpretation as the `K correction' from the observed $K_{\rm em}$ to
the true $K_2$ value \citep{munoz05} cannot be single valued.  Part of
the variation can be accounted for with time-dependent variations in
the disk shielding angle, $\alpha$, which are included in the models
of \citet{munoz05} but the \citet{bassa09} measurement of different
$K_{\rm em}$ values {\em simultaneously} in H$\alpha$ and He\,{\sc i}
must indicate different patterns of emission across the donor star.

\begin{figure*}
\includegraphics[width=12cm,angle=90]{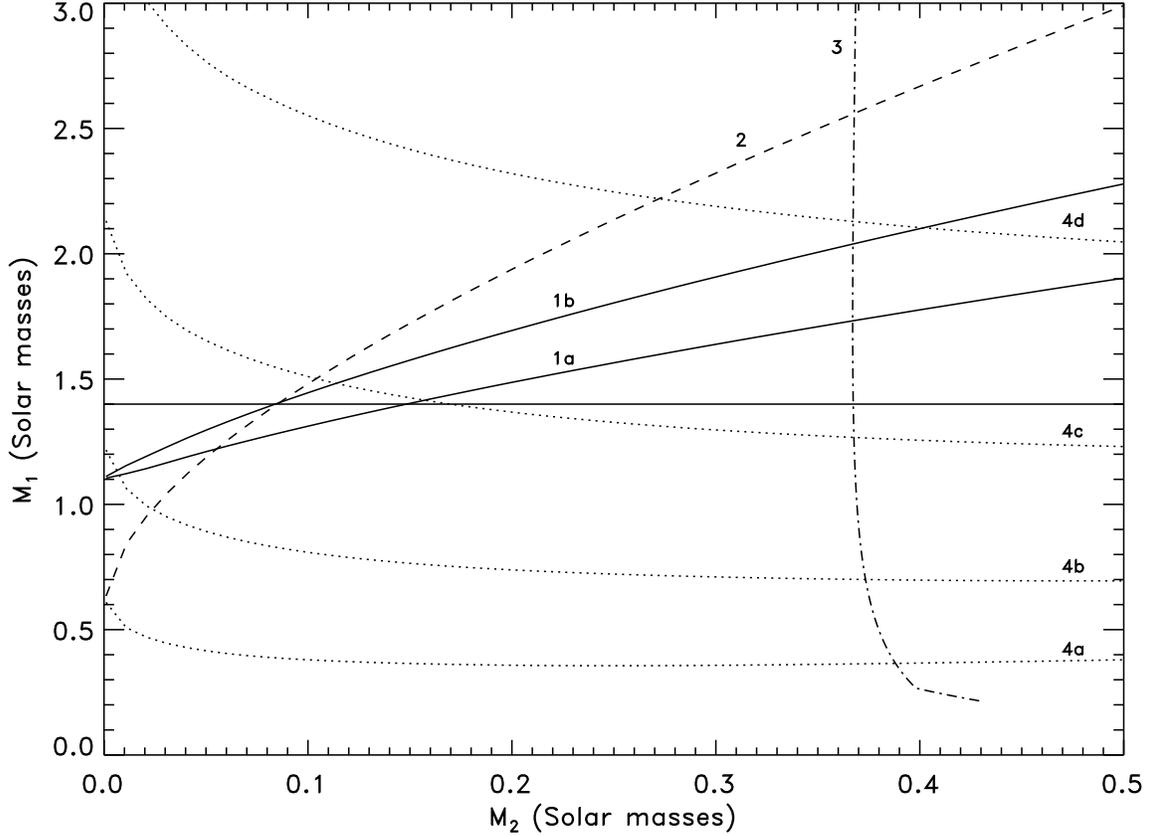}
\caption{Dynamical constraints on the mass of the neutron star ($M_1$)
  and the companion star ($M_2$).  The horizontal solid line
  corresponds to a canonical 1.4\,$_{\odot}$ neutron star.  The solid
  lines 1a and 1b are derived from the quiescent H$\alpha$ measurement
  of \citet{bassa09}.  The lower line, 1a is a lower boundary of
  allowed parameter space assuming only that $K_2 > K_{\rm em}$.  The
  higher line, 1b requires additionally that the H$\alpha$ emission
  originates in regions directly illuminated by the neutron star.  The
  dashed line, 2, is obtained from the He\,{\sc ii} measurement of
  \citet{munoz09}, assuming that $K_{\rm L_1} < K_{\rm em}$.
  The dot-dashed line 3 represents main-sequence donors that fill the
  companion star Roche lobe.  Finally, the dotted lines are
  constraints imposed by our measurements of the disk velocity.  Curves
  4a--d respectively correspond to outer disk velocities of 400, 500,
  600, and 700~km/s.}
\label{fig:params}
\end{figure*}

These limitations notwithstanding, we can attempt to constrain the
masses of the binary components using this information.  In some
respects we follow the discussion of \citet{munoz09} and
\citet{bassa09}, although we present the constraints more directly in
$M_2$ vs. $M_1$ space in Figure~\ref{fig:params}.  The first
constraint is that the $K_2$ value must be greater than all measured
$K_{\rm em}$.  The H$\alpha$ detection is most constraining in this
respect, requiring $K_2 \ga 405$\,km\,s$^{-1}$.  We show this
constraint in Figure~\ref{fig:params} as curve 1a which provides a
lower limit on $M_1$ as a function of $M_2$.  We can go beyond this
limit if we assume that the emission originates in excitation by
direct emission from near the compact object since the pole of the
companion star itself (corresponding to $K_{\rm em} = K_2$) is
shielded by the curve of the donor star's surface.  A tighter
constraint is that $K_{\rm em}$ is less than the velocity at the
tangent point, where rays from the compact object are tangential to
the stellar surface.  We show this constraint as curve 1b.

While there is uncertainty in the actual emission pattern over the
donor star, the center of light must be between the L$_1$ point and
the tangent point, so we also require that $K_{\rm L_1}$ be less than
the smallest measured $K_{\rm em}$, i.e.  \ $K_{\rm L_1} \la
310$\,km\,s$^{-1}$.  This constraint is shown as curve 2 in
Figure~\ref{fig:params} and this sets an upper limit on $M_1$ as a
function of $M_2$.  If $M_1$ lies above this line then the mass ratio
of the system is small and the donor is consequently too small to be
consistent with the range of velocities seen by different groups at
different times.

These constraints together define a band in $M_1-M_2$ space that
becomes increasingly narrow for small mass ratios.  This argument
(using curves 1a and 2) is essentially that used by \citet{bassa09} to
constrain $0.075<q<0.105$ ($0.1<M_2<0.15$) for the specific case of a
1.4\,$_{\odot}$ neutron star.  This case is barely consistent with
these constraints and requires an unusually extreme mass ratio and a
very low-mass donor.  Limiting emission to the tangent point rather
than the pole (the band defined by curves 1b and 2) virtually
eliminates a 1.4\,$_{\odot}$ neutron star from consideration. 

A 1.4\,M$_{\odot}$ neutron star is often assumed in LMXBs.  There is
some support for this in the observed tight distribution of masses of
radio pulsars in binaries, of $1.35\pm0.04$\,M$_{\odot}$
\citep{thorsett99}.  This tight distribution presumably reflects the
distribution in birth masses, however, whereas neutron stars in LMXBs
are accreting and so are likely to be of higher mass.  Indeed, we do
see rather larger masses where we can measure them in milli-second
pulsars which are believed to be the descendants of LMXBs
\citep{demorest10}.  Moving on to the companion star, the default
assumption is often that it is a main-sequence star, or at least lies
close to a main-sequence mass-radius relation.  While under-massive
mass donors are seen in some LMXBs, a main-sequence donor is still a
reasonable starting point in the absence of other evidence for a more
evolved status.  Such other evidence can sometimes be very strong, for
example, in abundance anomalies indicating that the donor has
experienced a prior period of CNO burning (e.g.\ XTE~J1118+480;
\citealt{haswell02}).  In the case of UY~Vol, however, as discussed in
Section~\ref{AbundanceSection}, it appears that the carbon abundance
is at least normal, if not enhanced.  If we follow the usual
prescription for deducing the mass of a main-sequence donor in an
interacting binary \citep{frank02} we obtain $M_2 =
0.42$\,M$_{\odot}$.  This is based on very simplistic assumptions
($R/R_{\odot} = M/M_{\odot}$ on the lower main-sequence), and more
realistic mass radius relations produce somewhat lower donor
masses. By way of example, we show in Figure~\ref{fig:params} the
mass-radius relation of \citet{patterson84} as curve 3.  This favors
masses around 0.37\,M$_{\odot}$ for plausible neutron star masses.

We can now draw one quite strong conclusion from the dynamical
evidence in Figure~\ref{fig:params}: if the donor star is on the
main-sequence, then the neutron star is massive.  If curve 1b is
obeyed (i.e.\ quiescent H$\alpha$ emission is driven by direct
irradiation from the neutron star or inner disk), then a main-sequence
companion star requires a neutron star more massive than about
2\,M$_{\odot}$.  This case would actually agree quite well with the
mass estimate of \citet{ozel2006} of $2.1\pm0.28$\,M$_{\odot}$.
Alternatively if we expect the neutron star to be closer to the
canonical 1.4\,M$_{\odot}$, then we require a significantly
under-massive donor.  In that case, we might expect to see evidence of
CNO processing as in, for example, XTE~J1118+480
\citep{haswell02}, and we clearly do not.  In this case, however,
the low mass may be a consequence of evaporation of an initially
sub-Solar donor in a `Black-Widow-like' scenario \citep{Ratti:2012a}
where no CNO processing has had timeto occur.

We have one final dynamical constraint that has not been included yet.
We observe the peak disk emission in He\,{\sc ii} around a projected
Keplerian disk velocity of $V_{\rm kep} \sin i = 500$\,km\,s$^{-1}$,
and it is quite well bounded by a range of 400--600\,km\,s$^{-1}$.
This turns out to be difficult to accommodate simultaneously with the
dynamical constraints from the donor star.  
For a given pair of $M_1,M_2$ values, the mass ratio and inclination
are uniquely determined (via X-ray eclipses), and so the projected
Keplerian velocity from the disk rim can be calculated.  We assume the
disk is tidally truncated at 0.9\,$R_{\rm lobe}$
\citep{Whitehurst:1991a}, and refer to this velocity as $V{\rm kep, tidal}$.
This should be the minimum allowed velocity in a Keplerian disk.  We
show in Figure~\ref{fig:params} a final set of curves (4a--d)
corresponding to the loci $V_{\rm kep, tidal} \sin i = 400$, 500, 600,
and 700\,km\,s$^{-1}$ respectively.  The 500\,km\,s$^{-1}$ curve is
not consistent with the dynamical constraints at all, and even
600\,km\,s$^{-1}$ is only consistent with the lowest $M_1$ and $M_2$
values, and requires that the He\,{\sc ii} emission be mildly
sub-Keplerian.  Accommodating a main-sequence donor would require that
the He\,{\sc ii} emission be sub-Keplerian by about 50\,\%.  The
dynamical constraints from the disk velocities thus favor relatively
small masses, but we cannot robustly rule out higher masses since all
plausible solutions require some degree of sub-Keplerian motion.

We note that there is a caveat to this last point.
\citet{Somero:2012a} have recently reported a similar analysis of
2A~1822-371, where they find He\,{\sc ii} emission with disk-like
morphology, but sub-Keplerian velocity.  In the case of 2A~1822-371
the parameters are better constrained, so this conclusion cannot be
avoided.  That object also appears to be accreting at a substantially
higher rate that UY~Vol, however, and with significant mass loss
\citep{Bayless:2010a,Burderi:2010a}.  \citet{Somero:2012a} argue that
their He\,{\sc ii} emission originates in a disk wind above the
orbital plane.  It is then quite likely that 2A~1822-371 is an
anomaly, and that UY~Vol (with lower mass transfer rate) should
exhibit more normal disk emission.

\section{Conclusions}

We have examined optical spectroscopy of the LMXB UY~Vol spanning
several orbits and observing nights.  Doppler tomography shows that
the strongest line emission, seen most characteristically in He\,{\sc
  ii}, originates in the vicinity of the disk rim downstream of the
stream-impact point.  This dominant component is observed to be quite
variable in intensity and in how far around the disk it extends.  The
variations are seen both within a night, and from night-to-night.
This component is superimposed on weaker, less variable emission from
the accretion disk.  The phasing of the variations is consistent with
the He\,{\sc ii} emission originating on the irradiated inner side of
a disk bulge region.  The higher excitation O\,{\sc VI} line is seen
at a higher velocity than He\,{\sc ii}, suggesting some stream
overflow.  It quite closely resembles the O\,{\sc v} line seen in the
far-UV by \citet{pearson06} in its kinematic behavior.

Careful examination of the Bowen blend reveals N\,{\sc iii} and
C\,{\sc iii} components moving in phase with the companion star.
These are best seen by Doppler correcting the line profiles using
radial velocity curves and they are maximized for $K_{\rm em}=
329\pm26$\,km/s, consistent with independent measurements of $K_{\rm
  em}$ by \citet{munoz09}, \citet{bassa09}, and \citet{Ratti:2012a}.
Combining these dynamical constraints with the inferred disk
velocities from our Doppler tomograms, we favor a neutron star close
to the canonical mass, $M_1\sim 1.5$\,M$_{\odot}$ and a very low mass
donor star, with $M_2\sim0.1$\,M$_{\odot}$, in spite of the lack of
evidence for CNO processing.  Even this solution requires that the
peak disk-emission be mildly sub-Keplerian.  Higher masses are
disfavored but cannot be securely ruled out.  A main-sequence donor
star ($M_2\simeq0.37$\,M$_{\odot}$) would require
$M_1>2.0$\,M$_{\odot}$ and peak disk emission at velocities 30\,\%\
lower than expected at the edge of the disk.

\acknowledgements We are grateful to Danny Steeghs for providing the
modulation mapping software, and for helpful discussions on our
results, and also to Tom Marsh for {\tt molly} and {\tt doppler}. This
work is supported by NASA/Louisiana Board of Regents grant NNX07AT62A/
LEQSF(2007-10) Phase3-02 and by the National Science Foundation under
Grant No. AST-0908789.

This work is based on observations obtained at the Gemini Observatory,
which is operated by the Association of Universities for Research in
Astronomy, Inc., under a cooperative agreement with the NSF on behalf
of the Gemini partnership: the National Science Foundation (United
States), the Science and Technology Facilities Council (United
Kingdom), the National Research Council (Canada), CONICYT (Chile), the
Australian Research Council (Australia), Minist\'{e}rio da Ci\^{e}ncia
e Tecnologia (Brazil) and Ministerio de Ciencia, Tecnolog\'{i}a e
Innovaci\'{o}n Productiva (Argentina). Our Gemini Program ID is
GS-2008A-Q-15. This research has also made use of NASA's Astrophysics
Data System.

%All my figures

\begin{deluxetable}{lllllll}
\tablecolumns{6}
\tablecaption{{\bf Log of GMOS Observations}} % from line_cent_new.out
\tablewidth{0pt}
\tablehead{\colhead{Date}  & \colhead{Start Time (MJD)} & \colhead{End Time (MJD)}  & \colhead{Start Phase} & \colhead{End Phase} & \colhead{Orbits}}
\startdata
\tableline
2008-02-09   &	54505.193	  &     54505.370 & 0.269 & 0.398 & 1.11\\
2008-02-10 & 	54506.103 & 54506.231  & 0.981 & 0.790 &  0.81\\
2008-02-11  &	54507.171 	  &	54507.337 	& 0.650 & 0.728 & 1.04	\\
2008-03-14  &	54539.019	 &	54539.049	& 0.576 & 0.772 & 0.19  \\
\tableline		 													  
\enddata
\label{tbl-1}
\end{deluxetable}

\begin{deluxetable}{lll}
\tablecolumns{3}
\tablecaption{{\bf IDENTIFIED LINES}} % from line_cent_new.out
\tablewidth{\textwidth}
\tablehead{\colhead{Line}  &\colhead{$\lambda_{air}$} & \colhead{$\lambda _c$ (\AA)} } 
\startdata
\tableline
H\,{\sc i} 2-6   & 4101.734 &  4101.6  $\pm$  0.6  \\
He\,{\sc ii} 4-12 & 4100.041 & \\
\tableline		 													  
  	      	       	     	   	       	     	   	     
He\,{\sc ii} 4-11    & 4199.832 &  4201. $\pm$  2. \\
\tableline		 													  

H\,{\sc i} 2-5  &  4340.464 & 4340.8  $\pm$   0.9  \\
He\,{\sc ii} 4-10 & 4338.671 & \\
\tableline		 													  
He\,{\sc ii} 4-9    &  4541.591 & -- \\
\tableline		 													  
  	      	       	     	   	       	     	   	     
C\,{\sc iii}  & 4647.4   &   4645.0   $\pm$   0.8  \\ 
       & 4650.3   &  \\
       & 4651.5   &  \\
N\,{\sc iii}  & 4634.2 & \\
	& 4640.6 & \\
	& 4641.9 & \\
\tableline		 													  

He \,{\sc ii} 3-4    &  4685.71 & 4686.3   $\pm$  0.8  \\
\tableline		 													  
  	      	       	     	   	       	     	   	     
H\,{\sc i} 2-4  & 4861.352  & 4861.4  $\pm$   0.7 \\
He \,{\sc ii} 4-8 & 4859.20 & \\
\tableline		 													  
  	      	       	     	   	       	     	   	     
He\,{\sc i}   &  4921.93 & 4922. $\pm$ 1\\ 
\tableline		 													  
  	      	       	     	   	       	     	   	     
He\,{\sc i}         &  5015.67 & 5015.3  $\pm$  0.8  \\
\tableline		 													  
C/N/O blend  &  -- & 5133  \\
\tableline		 													  
  	      	       	     	   	       	     	   	       	      	       	     	   	       	     	   	     
O\,{\sc vi}         &  5289. & 5289. $\pm$  1. \\
O\,{\sc vi}         &  5290. & . \\
\tableline		 													  
  	      	       	     	    	       	     	   	     
He \,{\sc ii} 4-7    & 5411.53 &  5412.  $\pm$  1.   \\

\tableline \enddata \tablecomments{Table of identified lines. The
  second column is the laboratory predicted central wavelength in
  air. The third column is the best estimate of the centroid as
  measured from the average spectra. Errors are estimated by creating
  two statistically independent spectra each summing over 34 of our
  nights. The $\lambda$4541 He\,{\sc ii} line falls on a chip gap, so
  no center is measured. The feature visible at 5133\AA\ is likely a
  blend of C/N/O.}
\label{tbl-2}
\end{deluxetable}

\end{document}